\documentclass[preprint,12pt]{elsarticle}
\journal{BioSystems}

\usepackage{graphicx}
\usepackage{multirow}
\usepackage{amsmath}
\usepackage{stfloats}

\usepackage{algorithm}
\usepackage{algorithmic}

\usepackage[english]{babel}
\usepackage[]{natbib}

\usepackage{url}
\makeatletter
\renewcommand\@biblabel[1]{#1} 

\newenvironment{proof}%
{\noindent\textit{Proof. }}%
{\textit{ Q.E.D.}}

\newtheorem{theorem}{Theorem}

\newtheorem{prop}{Proposition}

\def\blacksquare{\textit{QED}}

\def\calv{V'}

\def\cvt{\calv^{(t)}}

\def\ctz{C^{(0)}}

\def\mutt{\textbf{\textsf{M}}^{(t)}}
\def\mutz{\textbf{\textsf{M}}^{(0)}}
\def\muttp{\textbf{\textsf{M}}^{(t')}}

\def\muttl{\textbf{\textsf{M}}^{(t-1)}}

\def\sel{\textbf{\textsf{S}}^{(t)}}
\def\repl{\textbf{\textsf{R}}^{(t)}}

\def\deatht{\textbf{\textsf{D}}^{(t)}}
\def\birtht{\textbf{\textsf{B}}^{(t)}}

\def\Pr{\textbf{Pr}}
\def\EXif{\limt\EX^{(t)}_{\{i\}}}

\def\EX{\textbf{Ex}}

\def\EXs{\EX^{(t)}_C}

\def\EXsl{\EX^{(t-1)}_C}

\def\lim{\mathit{lim}}
\def\limt{\mathop{\lim}_{t\rightarrow\infty}}

\def\calp{\mathcal{P}}

\def\dini{{k_{\textit{in}}^{(i)}}}

\def\douti{{k_{\textit{out}}^{(i)}}}

\def\I{\textbf{I}}
\def\incchi{\chi_{{+}}}
\def\decchi{\chi_{{-}}}
\def\incchir{\chi^{(r)}_{{+}}}
\def\decchir{\chi^{(r)}_{{-}}}
\def\incchirp{\chi^{(r')}_{{+}}}
\def\decchirp{\chi^{(r')}_{{-}}}
\def\incchio{\chi^{(1)}_{{+}}}
\def\decchio{\chi^{(1)}_{{-}}}

\def\incchin{\chi^{(\mathit{nd})}_{{+}}}
\def\decchin{\chi^{(\mathit{nd})}_{{-}}}
\def\inc{\textit{inc}}
\def\dec{\textit{dec}}
\def\fPr{F^{(r)}}
\def\fPo{F^{(1)}}
\def\fPrp{F^{(r')}}

\def\PCt{F_{t|C}}
\def\PCtl{F_{t-1|C}}
\def\tcv{t_{convg}}
\def\tfx{t_{fix}}
\def\muttcv{\textbf{\textsf{M}}^{(\tcv)}}
\def\prmnt{P_{\min,t}}
\def\prmntl{P_{\min,t-1}}
\def\prdmnt{\Pr(\Delta\mutt_{\min})}
\def\prdmntp{\Pr(\Delta\textbf{\textsf{M}}^{(t')}_{\min})}
\def\calpttp{\theta^{(t)}_{t'}}
\def\calpptt{\theta^{(t')}_{t}}

\def\calpt{\mathcal{P}_C^{(t)}}
\def\calptp{\mathcal{P}_C^{(t')}}

\begin{document}
\begin{frontmatter}

\title{A Novel Analytical Method for Evolutionary Graph Theory Problems}

\author{Paulo Shakarian}
\address{Network Science Center and Dept. of Electrical Engineering and Computer Science, United States Military Academy, West Point, NY 10996}

\author{Patrick Roos}
\address{Dept. of Computer Science, University of Maryland, College Park, MD 20740}

\author{Geoffrey Moores}
\address{Network Science Center and Dept. of Electrical Engineering and Computer Science, United States Military Academy, West Point, NY 10996}

\label{firstpage}

\begin{abstract}
Evolutionary graph theory studies the evolutionary dynamics of populations structured on graphs. A central problem is determining the probability that a small number of mutants overtake a population. Currently, Monte Carlo simulations are used for estimating such fixation probabilities on general directed graphs, since no good analytical methods exist. In this paper, we introduce a novel deterministic framework for computing fixation probabilities for strongly connected, directed, weighted evolutionary graphs under neutral drift. We show how this framework can also be used to calculate the expected number of mutants at a given time step (even if we relax the assumption that the graph is strongly connected), how it can extend to other related models (e.g. voter model), how our framework can provide non-trivial bounds for fixation probability in the case of an advantageous mutant, and how it can be used to find a non-trivial lower bound on the mean time to fixation. We provide various experimental results determining fixation probabilities and expected number of mutants on different graphs. Among these, we show that our method consistently outperforms Monte Carlo simulations in speed by several orders of magnitude. Finally we show how our approach can provide insight into synaptic competition in neurology.
\end{abstract}
\end{frontmatter}

\section{Introduction}
Evolutionary graph theory (EGT), introduced by ~\cite{lieberman_evolutionary_2005}, studies the problems related to population dynamics when the underlying structure of the population is represented as a directed, weighted graph.  This model has been applied to problems in evolutionary biology \citep{zhang07}, physics \citep{sood08}, game theory \citep{pacheco06}, neurology \citep{turney12}, and distributes systems \citep{jiang12}.  A central problem in this research area is computing the \textit{fixation probability} - the probability that a certain subset of mutants overtakes the population.  Although good analytical approximations are available for the undirected/unweighted case \citep{antal06,Broom2010}, these break down for directed, weighted graphs as shown by \cite{masuda09}.  As a result, most work dealing with evolutionary graphs rely on Monte Carlo simulations to approximate the fixation probability \citep{rychtar08,broom08,barbosa2010}.  In this paper we develop a novel deterministic framework to compute fixation probability in the case of neutral drift (when mutants and residents have equal fitness) in directed, weighted evolutionary graphs based on the convergence of ``vertex probabilities'' to the fixation probability as time approaches infinity.  We then show how this framework can be used to calculate the expected number of mutants at a given time, how the framework can be modified to do the same for related models, how it can provide non-trivial bounds for fixation probability in the case of an advantageous mutant, and how it can provide a non-trivial lower bound on the mean time to fixation.  We also provide various experiments that show how our method can outperform Monte Carlo simulations by several orders of magnitude.  Additionally, we show that the results of this paper can provide direct insight into the problem of synaptic competition in neurology.

Our method also fills a few holes in the literature.  First, it allows for deterministic computation of fixation probability when there is an initial \textit{set} of mutants --  not just a singleton (the majority of current research on evolutionary graph theory only considers singletons).  Second, it allows us to study how the mutant population changes as a function of time.  Third, we show (by way of rigorous proof) that fixation probability, under the case of neutral drift is a lower bound for the case of the advantageous mutant - confirming simulation observations by \cite{masuda09a}.  Fourth, we show (also by way of rigorous proof) that fixation probability under neutral drift is additive (even for weighted, directed graphs), which extends the work of ~\cite{Broom2010} that proved this for undirected/unweighted graphs.  Fifth, we provide a non-trivial lower bound for the computation of mean time to fixation in the general case - which has only previously been explored for well-mixed populations~\citep{antalTime06} and special cases of graphs~\citep{broom09speed}.

This paper is organized as follows.  In Section~\ref{fix-prob} we review the original model of Lieberman et al., introduce the idea of ``vertex probabilities'' and show how they can be used to find the fixation probability.  We then show how this can be used to determine the expected number of mutants at a given time in Section~\ref{exmut-sec}.  This is followed by a discussion of how the framework can be extended to other update rules in Section~\ref{altUpdRules} and then for bounding fixation probability in the case of an advantageous mutant in Section~\ref{advMutBound}.  We then discuss how our approach can be adopted to bound mean time to fixation in Section~\ref{ttf-sec}.  We use the results of the previous sections to create an algorithm for computing fixation probability and introduce a heuristic technique to significantly decrease the run-time.  The algorithm and several experimental evaluations are described in Section~\ref{exper}.  In Section~\ref{appSec}, we show how our framework can be applied to neurology to gain insights into synaptic competition. Finally, we discuss related work in Section~\ref{rw-sec} and conclude.

\section{Directly Calculating Fixation Probability}
\label{fix-prob}

The classic evolutionary model known as the \textit{Moran Process} is a stochastic process used to describe evolution in a well-mixed population~\citep{moran58}.  All the individuals in the population are either \textit{mutants} or \textit{residents}.  The aim of such work was to determine if a set of mutants could take over a population of residents (achieving ``fixation'').  In ~\cite{lieberman_evolutionary_2005}, evolutionary graph theory (EGT) is introduced, which generalizes the model of the Moran Process by specifying relationships between the $N$ individuals of the population in the form of a directed, weighted graph.  Here, the graph will be specified in the usual way as $G=(V,E)$ where $V$ is a set of nodes (individuals) and $E \subseteq V \times V$.  In most literature on evolutionary graph theory, the evolutionary graph is assumed to be \textit{strongly connected}.  We make the same assumption and state when it can be relaxed.

For any node $i$, the numbers $\dini,\douti$ are the in- and out- degrees respectively.  We will use the symbol $N$ to denote the sized of $V$.  Additionally, we will specify weights on the edges in set $E$ using a square matrix denoted $W=[w_{ij}]$ whose side is of size $N$.  Intuitively, $w_{ij}$ is the probability that member of the population $j$ is replaced by $i$ given that member $i$ is selected.  We require $\sum_{j}w_{ij}=1$ and that $(i,j)\in E$ iff $w_{ij} > 0$.  If for all $i,j$, we have $w_{ij}=1/\douti$, then the graph is said to be ``unweighted.''  If for all $(i,j)\in E$, we have $(j,i) \in E$ the graph is said to be ``undirected.''  Though our results primarily focus on the general case, we will often refer to the special case of undirected/unweighted graphs as this special case is quite common in the literature~\citep{antal06,Broom2010}.

In this paper we will often consider the outcome of the evolutionary process when there is a set of initial mutants as opposed to a singleton.  Hence, we say some set (often denoted $C$) is a \textit{configuration} if that set specifies the set of mutants in the population (all other members in the population then are residents).  We assume all members in the population are either mutants or residents and have a fitness specified by a parameter $r>0$.  Mutants have a fitness $r$ and residents have a fitness of $1$.  At each time step, some individual $i \in V$ is selected for ``birth'' with a probability proportional to its fitness.  Then, an outgoing neighbor $j$ of $i$ is selected with probability $w_{ij}$ and replaced by a clone of $i$.  Note if $r=1$, we say we are in the special case of \textit{neutral drift.}   

We will use the notation $P_{\calv,t}$ to refer to the probability of being in configuration $V'$ after $t$ timesteps and $P_{V',t|C}$ to be the probability of being in configuration $V'$ at time $t$ conditioned upon initial configuration $C$.  Perhaps the most widely studied problem in evolutionary graph theory is to determine the fixation probability.  Given set of mutants $C$ at time $0$, the fixation probabilty is defined as follows.
\begin{equation}
F_C=\lim_{t \rightarrow \infty}P_{V,t|C}
\end{equation}

\noindent This is the probability that an initial set $C$ of mutants takes over the entire population as time approaches infinity.  Similarly, we will use the term the \textit{extinction probability}, $\overline {F_C}$, to be $\lim_{t \rightarrow \infty}P_{\emptyset, t|C}$.  If the graph if strongly connected, then we have $F_C + \overline {F_C} =1$.  Hence, for a strongly connected graph, a mutant either fixates or becomes extinct.  Typically, this problem is studied using Monte Carlo simulation.  This work uses the idea of a \textit{vertex probabilities} to create an alternative to such an approach.  The vertex probability is the probability that a certain vertex is a mutant at a certain time given an initial configuration.  For vertex $i$ at time $t$, we denote this as  $P_{i,t|C}$.  Often, for ease of notation, we shall assume that the probabilities are conditioned on some initial configuration and drop the condition, writing $P_{i,t}$ instead of $P_{i,t|C}$.  We note that $P_{i,t}$ can be expressed in terms of probabilities of configurations as follows.
\begin{equation}
P_{i,t}=\mathop{\sum_{V' \in 2^V}}_{\textit{s.t. }i \in V'}P_{V',t}
\end{equation}

\noindent Viewing the probability that a specific vertex is a mutant at a given time has not, to our knowledge, been studied before with respect to evolutionary graph theory (or in related processes such as the voter model).  The key insight of this paper is that studying these probabilities sheds new light on the problem of calculating fixation probabilities in addition to providing other insights into EGT.  For example, it is easy to show the following relationship.

\begin{prop}
\label{init-set-prop}
Let $\calv$ be a subset of $V$ and $t$ be an arbitrary time point.  Iff for all $i \in \calv$, $P_{i,t}=1$ and for all $i \notin \calv$, $P_{i,t}=0$, then $P_{\calv,t}=1$ and for all $\calv' \in 2^V$ s.t. $\calv' \not\equiv \calv$, $P_{\calv',t}=0$.
\end{prop}

It is easy to verify that $F_{C} > 0$ iff $\forall i \in V$, $\lim_{t \rightarrow \infty}P_{i,t}>0$.  Hence, in this paper, we shall generally assume that $\lim_{t \rightarrow \infty}P_{i,t}>0$ holds for all vertices $i$ and specifically state when it does not.  As an aside, for a given graph, this assumption can be easily checked: simply ensure for $j \in V-C$ that exists some $i \in C$ s.t. there is a directed path from $i$ to $j$.	 
	      
Now that we have introduced the model and the idea of vertex probabilities we will show how to leverage this information to compute fixation probability.  It is easy to show that as time approaches infinity, the vertex probabilities for all vertices converge to the fixation probability when the graph if strongly connected.  

\begin{theorem}
\label{eq-prbs}
$\forall i,  \limt P_{i,t|C} = F_C$
\end{theorem}
\noindent Now let us consider how to calculate $P_{i,t}$ for some $i$ and $t$.  For $t=0$, where we know that we are in the state where only vertices in a given set are mutants, we need only appeal to Proposition~\ref{init-set-prop} - which tells us that we assign a probability of $1$ to all elements in that set and $0$ otherwise.  For subsequent timesteps, we have developed Theorem~\ref{prob-thm} shown next (the proof of which is included in the supplement).

\begin{theorem}
\label{prob-thm}
\begin{eqnarray*}
P_{i,t}&=&P_{i,t-1}+\mathop{\sum}_{(j,i)\in E}w_{ji} \left( P_{j,t-1}\cdot S_{(j,t)|(j,t-1)}-P_{i,t-1}\cdot S_{(j,t)|(i,t-1)} \right)
\end{eqnarray*}
\noindent ($S_{(j,t)|(i,t-1)}$ is the probability that $j$ is picked for reporduction at time $t$ given that $i$ was a mutant at time $t-1$.)
\end{theorem}

\noindent We believe that a concise, tractable analytical solution for $S_{(j,t)|(i,t-1)}$ is unlikely.  However, for neutral drift ($r=1$), these conditional probabilities are trivial - specifically, we have for all $i,j,t$, $S_{(j,t)|(i,t-1)} = 1/N$ as this probability of selection is independent of the current set of mutants or residents in the graph.  Hence, in the case of neutral drift, we have the following:

\begin{eqnarray}
\label{nd-eqn}
P_{i,t}&=&P_{i,t-1}+\sum_{(j,i)\in E}\frac{w_{ji}}{N}\cdot \left(P_{j,t-1}-P_{i,t-1}\right)
\end{eqnarray}

\noindent Studying evolutionary graph theory under neutral drift was a central theme in several papers on EGT in the past few years~\citep{Broom2010,masuda09a} as it provides an intuition on the effects of network topology on mutant spread.  In Section~\ref{advMutBound} we examine the case of the advantageous mutant ($r>1$).  Neutral drift allows us to strengthen the statement of Equation~\ref{eq-prbs} to a necessary and sufficient condition - showing that when the probabilities of all nodes are equal, then we can determine the fixation probability.

\begin{theorem}
\label{convg-a}
Assuming neutral drift ($r=1$), given initial configuration $C$ with fixation probability $F_C$, if at time $t$ the quantities $P_{i,t|C}$ are equal (for all $i \in V$), then they also equal $F_C$.
\end{theorem}

\noindent Therefore, under neutral drift, we can determine fixation probability when Equation~\ref{nd-eqn} causes all $P_{i,t}$'s to be equal.  We can also use Equation~\ref{nd-eqn} to find bounds on the fixation probability for some time $t$ by the following result that holds for any time $t$ under neutral drift.

\begin{equation}
\label{nd-bnd-thm}
\min_i P_{i,t} \leq F_C \leq \max_i P_{i,t}
\end{equation}

Under neutral drift, we can show that fixation probability is additive for disjoint sets.  Broom et al. proved a similar result the a special case of undirected/unweighted evolutionary graphs~\cite{Broom2010}.  However, our proof (contained in the supplement) differs from theirs in that we leverage Equation~\ref{nd-eqn}.  Further, unlike the result of Broom et al., our result applies to the more general case of weighted, directed graphs.

\begin{theorem}
\label{add-thm}
When $r=1$ for disjoint sets $C,D \subseteq V$, $F_C+F_D = F_{C \cup D}$.
\end{theorem}

\section{Calculating the Expected Number of Mutants}
\label{exmut-sec}
In addition to allowing for the calculation of fixation probability, our framework can also be used to observe how the expected number of mutants changes over time.  We will use the notation $\EXs$ to denote the expected number of mutants at time $t$ given initial set $C$.  Formally, this is defined below.

\begin{equation}
\EXs = \sum_{i \in V}P_{i,t}
\end{equation}

Unlike fixation probability, which only considers the probability that mutants overtake a population, $\EXs$ provides a probabilistic average of the number of mutants in the population under a finite time horizon.  For example, is has been noted that graph structures which amplify fixation normally also increase time to absorption~\citep{broom09a,paley07}.  Hence, finding the expected number of mutants may be a more viable topic in some areas of research where time is known to be limited.  Following from Equation~\ref{nd-eqn} where we showed how to compute $P_{i,t}$ for each node at a given time, we have the following relationship concerning the expected number of mutants at a given time under neutral drift.

\begin{eqnarray}
\label{nd-cor2}
\EXs &=& \EXsl+\frac{\EXsl}{N}-\frac{1}{N}\sum_{i \in V}\sum_{(j,i)\in E}w_{ji}\cdot P_{i,t-1}
\end{eqnarray}

Based on Equation~\ref{nd-cor2}, we notice that for $r=1$, at each time-step, the number of expected mutants increases by at most the average fixation probability and decreases by a quantity related to the average ``temperature.''  The temperature of vertex $i$ (denoted $T_i$) is defined for a given node is the sum of the incoming edge weights~\cite{lieberman_evolutionary_2005}: $T_i = \sum_j w_{ji}$.  Intuitively, nodes with a higher temperature change more often between being a mutant and being a resident than those with lower temperature.  Re-writing Equation~\ref{nd-cor2} in terms of temperature we have the following:

\begin{eqnarray}
\label{nd-cor2rewrite}
\EXs &=& \EXsl+\frac{\EXsl}{N}-\frac{1}{N}\sum_{i \in V}T_i\cdot P_{i,t-1}
\end{eqnarray}

Hence, if the preponderance of high temperature nodes are likely to be mutants, then most likely the average number of mutants will decrease at the next time step.  We also note that Theorem~\ref{prob-thm}, Equation~\ref{nd-eqn}, and Equation~\ref{nd-cor2} do not depend on the assumption that the underlying graph is strongly connected.  Therefore, as such is the case, we can study the relationship of time vs. expected number of mutants for \textit{any} evolutionary graph (under neutral drift).  This could be of particular interest to non-strongly connected evolutionary graphs that may have trivial fixation probabilities (i.e. $1$ or $0$) but may have varying levels of mutants before achieving an absorbing state.

\section{Applying the Framework to Other Update Rules}
\label{altUpdRules}

The results of the last two sections not only apply to the original model of ~\cite{lieberman_evolutionary_2005}, but several other related models in the literature.  Viewing an evolutionary graph problem as a stochastic process, where the states represent different mutant-resident configurations, it is apparent that the original model specifies the transition probabilities.  However, there are other ways to specify the transition probabilities known as update rules.  Several works address different update rules~\citep{antal06,sood08,masuda09a}.  Overall, we have identified three major families of update rules - birth-death (a.k.a. the invasion process) where the node to reproduce is chosen first, death-birth (a.k.a. the voter model) where the node to die is chosen first, and link dynamics, where an edge is chosen.  We summarize these in Table~\ref{fig:moran-var-chart-a}.

\begin{table}
\caption{Different families of update rules.}
\label{fig:moran-var-chart-a}
\begin{center}
\begin{tabular}{|l|l|}
\hline
Update Rule  & Intuition\\
\hline
\hline
Birth-Death (BD)& (1) Node $i$ selected \\
(a.k.a. Invasion Process (IP)) & (2) neighbor of $i$, node $j$ selected \\
 & (3) Offspring of $i$ replaces $j$\\
\hline
Death-Birth (DB)& (1) Node $i$ selected \\
(a.k.a. Voter Model (\textsf{VM})) & (2) neighbor of $i$, node $j$ selected \\
 & (3) Offspring of $j$ replaces $i$\\
\hline
Link Dynamics (LD)& (1) Edge $(i,j)$ selected \\
 & (2) The offspring of one node in the \\
 & edge replaces the other node \\

\hline
\hline
\end{tabular}
\end{center}
\end{table}

We have already shown how our methods can deal with the original model of Lieberman et al., often referred to as the Birth-Death (BD) process.  In this section, we apply our methods to the neutral-drift (non-biased) cases of death-birth and link-dynamics. In these models, the weights of the edges is typically not considered.  Hence, in order to align this work with the majority of literature on those models, we will express vertex probabilities in terms of node in-degree ($\dini$) and the set of directed edges ($E$).  We note that these results can be easily extended to a more general case with an edge-weight matrix as we used for the original model of EGT.

\subsection{Death-Birth Updating}
\label{dbSec}
Under the death birth model (DB), at each time step, a vertex $i$ is selected for death.  With a death-bias (DB-D), it is selected proportional to the inverse of its fitness, with a birth-bias (DB-B) it is selected with a probability $1/N$, which is also the probability under neutral drift.  Then, an incoming neighbor ($j$) is selected either proportional to the fitness of all incoming neighbors (birth-bias), or with a uniform probability (in the case death-bias or neutral drift).  The selected neighbor then replaces $i$.  Here, we compute $P_{i,t}$ under this dynamic with $r=1$.

\begin{eqnarray}
\label{db}
P_{i,t}=( 1-N^{-1}) P_{i,t-1}+(N \dini)^{-1}\mathop{\sum}_{(j,i)\in E} P_{j,t-1}
\end{eqnarray}

We note that the proof of convergence still holds for death-birth - that is for some time $t$, $\forall i$, the value $P_{i,t}$ is the same, then $P_{i,t}=F_C$.  Further, Theorem~\ref{add-thm} holds for DB under neutral drift as well, specifically, for disjoint sets $C,D \subseteq V$, $F_C+P_D = P_{C \cup D}$.

\subsection{Link-Dynamics}
With link dynamics (LD), at each time step an edge $(i,j)$ is selected either proportional to the fitness of $i$ or the inverse of the fitness of $j$.  It has previously been shown that LD under birth bias is an equivalent process to LD with a death bias~\cite{masuda09a}.  Under neutral drift, the probability of edge selection is $1/|E|$ (where $|E|$ is the cardinality of set $E$).  Then, $i$ replaces $j$.  Now, we compute $P_{i,t}$ under this dynamic with $r=1$.
\begin{eqnarray}
\label{ld}
P_{i,t}=( 1-\dini|E|^{-1} ) P_{i,t-1}+\frac{1}{|E|}\mathop{\sum}_{(j,i) \in E}P_{j,t-1}
\end{eqnarray}

Again, convergence and additivity of the fixation probability still hold under link dynamics just as with BD and DB.

\section{Bounding Fixation Probability for $r>1$}
\label{advMutBound}

So far we have shown how our method can be used to find fixation probabilities under the case of neutral drift.  Here, we show how our framework can be useful in the case of an advantageous mutant (when the value for $r$, the relative fitness, is greater than $1$).  First, we show that our method provides a lower bound.  We then provide an upper bound on the fixation probability that can be used in conjunction with our framework when studying the case of the advantageous mutant.  We note that certain parts of these proofs are specific for diffent update rules, and we identify them using the abbreviations from the last section (DB-D, DB-B, and LD).  The update of the original model of \cite{lieberman_evolutionary_2005} is known as the ``birth-death'' model and abbreviated BD.  If the fitness bias is on a birth event, we denote it as BD-B and if the bias is on a death event we denote it as BD-D.

Naoki Masuda observes experimentally (through simulation) that the fixation probability computed with neutral drift appears to be a lower bound on the fixation probability for an advantageous mutant~\citep{masuda09a}.  We were able to prove this result analytically -- the proof is included in the supplementary materials.

\begin{theorem}
\label{bd-mono}
For a given set $C$, let $\fPo_C$ be the fixation probability under neutral drift and $\fPr_C$ be the fixation probability calculated using a mutant fitness $r>1$.  Then, under BD-B, BD-D, DB-B, DB-D, or LD dynamics, $\fPo_C \leq \fPr_C$.
\end{theorem}

This proof leads to the conjecture that $r'>r$ implies $\fPrp_C \geq \fPr_C$. However, we suspect that proving this monotonicity property will require a different technique than used in Theorem~\ref{bd-mono}.  Next, to find an upper bound that corresponds with the lower bound above, we use the proof technique introduced in \cite{egtFpras}, to obtain the following non-trivial upper bounds of fixation probability for individual nodes in various update rules.

\begin{eqnarray}
\mathbf{BD-B:} & F_{\{i\}}\leq & r (r+\sum_j w_{ji})^{-1}\\
\mathbf{BD-D:} & F_{\{i\}}\leq & \left( {\sum_j \frac{w_{ji}}{r-rw_{ji}+w_{ji}}} \right)^{-1}\\
\mathbf{DB-B:} & F_{\{i\}}\leq &\sum_j r w_{ij}(1-w_{ij}+r w_{ij})^{-1}\\
\mathbf{DB-D:} & F_{\{i\}}\leq & r \sum_j w_{ij}
\end{eqnarray}

\section{A Lower Bound for Mean Time to Fixation}
\label{ttf-sec}
Another important, although less-studied problem with respect to evolutionary graph theory is the mean time to fixation - the average time it takes for a mutant to take over the population.  Closely related to this problem are mean time to extinction (average time for the resident to take over) and mean time to absorption (average time for either mutant or resident to take over).  This has been previously studied under the original Moran process for well mixed populations~\citep{antalTime06} as well as some special cases of graphs~\citep{broom09speed}.  However, to our knowledge, a general method to compute these quantities (without resorting to the use of simulation) have not been previously studied.  Here we take a ``first step'' toward developing such a method by showing how the techniques introduced in this paper can be used to compute a non-trivial lower bound for mean time to fixation (and easily modified to bound mean time to extinction and absorption).

Let $\PCt$ be the probability of fixation at time $t$.  Therefore, $\PCt-\PCtl$ is the probability of entering fixation at time $t$.  The symbol $t_C$ is the mean time.  By the results of \cite{antalTime06}, we have the following:

\begin{theorem}
\begin{eqnarray*}
t_C = \frac{1}{F_C}\sum_{t=1}^{\infty}t\cdot (\PCt-\PCtl)
\end{eqnarray*}
\end{theorem}

Our key intuition is noticing that at each time step $t$, $\PCt \leq\min_i P_{i,t}$.  From this, we use the accounting method to provide a rigorous proof for the following theorem that provides a non-trivial lower-bound for the mean time to fixation.  This result can be easily modified for mean time to extinction and absorption as well.
\begin{theorem}
\label{timeThm}
$\frac{1}{F_C}\sum_{t=1}^{\infty}t\cdot (\prmnt-\prmntl)\leq \frac{1}{F_C}\sum_{t=1}^{\infty}t\cdot (\PCt-\PCtl)$
Where $\prmnt = \min_i P_{i,t}$.
\end{theorem}

\section{Algorithm and Experimental Evaluation}
\label{exper}
\noindent We leverage the finding of the previous sections in Algorithm~\ref{dirsolutionalg}. As described earlier, our method has found the exact fixation probability when all the probabilities in $\bigcup_i\{ P_{i,t} \}$ (represented in the pseudo-code as the vector $\mathbf p$) are equal. We use Equation~\ref{nd-bnd-thm} to provide a convergence criteria based on value $\epsilon$, which we can prove to be the tolerance for the fixation probability.

\begin{prop}
\label{alg-guar}
Algorithm~\ref{dirsolutionalg} returns the fixation probability $F_C$ within $\pm\epsilon$.
\end{prop}

\begin{algorithm}
\caption{ - Our Novel Solution Method to Compute Fixation Probabilities
\newline
 {\sf {\bf Input}: Evolutionary Graph $\langle N, V, W \rangle$, configuration $C \subseteq V$, natural number $R > 0$, and real number $\epsilon\geq 0$.  \newline
{\bf Output}: Estimate of fixation probability of mutant.}}
\label{dirsolutionalg}
\vspace{2mm}
\begin{algorithmic}[1]
\STATE $p_i$ is the $i$th position in vector $p$ corresponding with vertex $i \in V$.
\STATE Set $p_i=1$ if $i \in C$ and $p_i=0$ otherwise. \COMMENT{As per Proposition~\ref{init-set-prop}}
\STATE ${\mathbf q}\gets{\mathbf p}$  \COMMENT{${\mathbf q}$ will be ${\mathbf p}$ from the previous time step.}
\STATE $\tau \gets 1$
\WHILE{$\tau > \epsilon$}
	\FOR{$i \in V$ \COMMENT{This loop carries out the calculation as per Equation~\ref{nd-eqn}}} 
		\STATE $sum\gets0$
		\STATE ${\mathbf m} \gets \{j \in V | w_{ji} > 0\}$
		\FOR{$j \in m$}
			  \STATE $sum =  sum + w_{ji} \cdot ({\mathbf q}_{j} - {\mathbf q}_{i})$ 
		\ENDFOR
		\STATE ${\mathbf p}_{i} \gets {\mathbf q}_{i} + 1/N \cdot sum$
	\ENDFOR
	\STATE ${\mathbf q}\gets{\mathbf p}$
	\STATE\label{tau-step} $\tau \gets (1/2)\cdot(\max p - \min p)$ \COMMENT{Ensures error bound based on Equation~\ref{nd-bnd-thm}}
\ENDWHILE
\STATE\label{ret-step} \textbf{return} $(\min p)+\tau$
\end{algorithmic}
\end{algorithm}

Our novel method for computing fixation probabilities on strongly connected directed graphs allows us to compute near-exact fixation probabilities within a desired tolerance. The running time of the algorithm is highly dependent on how fast the vertex probabilities converge. In this section we experimentally evaluate how the vertex probabilities in our algorithms converge. We also provide results from comparison experiments to support the claim that  \textit{Algorithm~\ref{dirsolutionalg}-ACC} finds adequate fixation probabilities order of magnitudes faster than Monte Carlo simulations.  We also show how the algorithm can be used to study the expected number of mutants as well as bound mean time to fixation.

\begin{figure}[t]
  \centering
  \includegraphics[width=1\textwidth]%
    {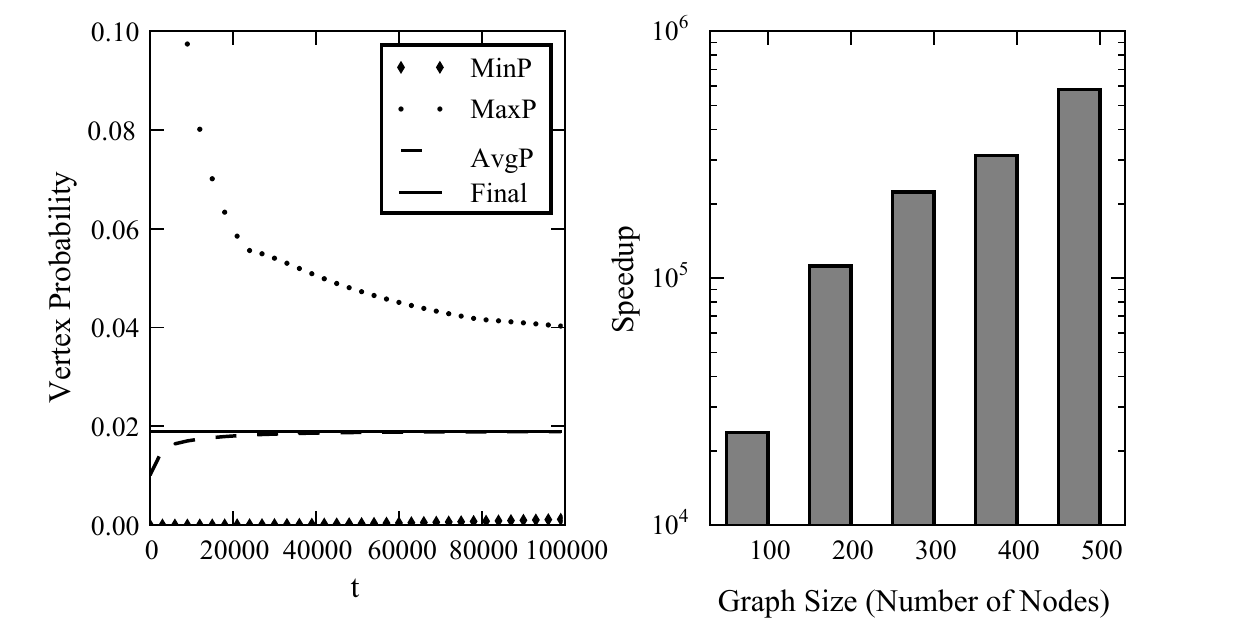}
      \caption{Left: Convergence of the minimum (MinP), maximum (MaxP), and average (AvgP) of vertex probabilities towards the final fixation probability as a function of our algorithm's iterations t for a graph of 100 nodes.  Right: Average speedup (on a log scale) for finding fixation probabilities achieved by our algorithm vs Monte Carlo simulation for graphs of different sizes.} \label{minmaxavg}
\end{figure}

\begin{figure}[ht]
  \centering
  \includegraphics[width=1\textwidth]%
    {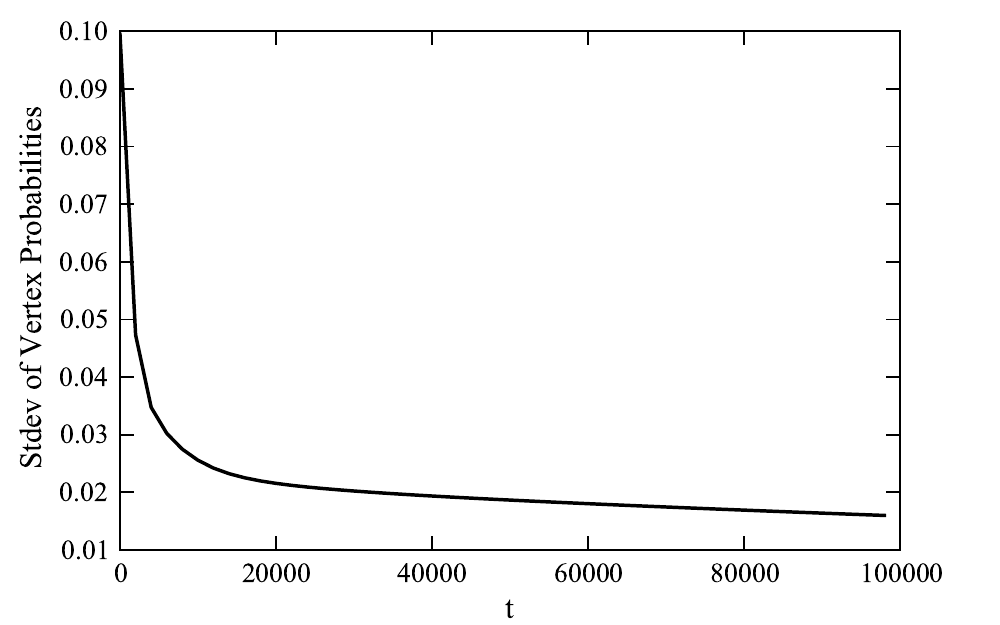}
  \caption{Standard deviation of vertex probabilities as a function of our algorithm's iterations for the same 100 node graph of Figure~\ref{minmaxavg} (left).} \label{stdev}
\end{figure}

\subsection{Convergence of Vertex Probabilities}

We ran our algorithm to compute fixation probabilities on randomly weighted and strongly connected directed graphs in order to experimentally evaluate the convergence of the vertex probabilities. We generated the graphs to be scale-free using the standard preferential attachment growth model~\citep{barabasi} and randomly assigned an initial mutant node. We replaced all edges in the graph given by the growth model with two directed edges and then randomly assigned weights to all the edges.

To compare Algorithm~\ref{dirsolutionalg} with the Monte Carlo approach, we should set the parameter $R$ in that algorithm to be comparable with $\epsilon$ in Algorithm~\ref{dirsolutionalg}.  As $\epsilon$ is the provable error of a solution to Algorithm~\ref{dirsolutionalg}.  Based on the commonly-accepted definition of estimated standard error from statistics, we can obtain the estimated standard error for the solution returned by Monte Carlo approach with the following expression (where $R$ is the number of simulation runs).

\begin{equation}
\label{std-err-prop}
\sqrt{\frac{F_C(1-F_C)}{R-1}}
\end{equation}

We can use Equation~\ref{std-err-prop} to estimate the parameter $R$ for the Monte Carlo approach as follows.  We set $\epsilon$ equal to the estimated standard error as per Expression~\ref{std-err-prop} and manipulate it algebraically.  This gives us $R\approx\frac{S(S-1)}{\epsilon^2}+1$ where $S$ is the solution to Algorithm~\ref{dirsolutionalg}, $\epsilon$ is the input parameter for Algorithm~\ref{dirsolutionalg} and $R$ is the number of simulation runs in the Monte Carlo approach that we estimate to provide a comparable error bound.  We also note, that as the vertex probabilities converge, the standard deviation of the $p$ vector in Algorithm~\ref{dirsolutionalg} could be a potentially faster convergence criteria. Note that using standard deviation of $p$ and returning the average vertex probability would no longer provide us of the guarantee in Proposition~\ref{alg-guar}, however it may provide good results in practice.  The modifications to the algorithm would be as follows: line~\ref{tau-step} would be  $\tau \gets \textsf{st.dev}(p)$ and line~\ref{ret-step} would be \textbf{return} $\textsf{avg}(p)$.  We will refer to this as \textit{Algorithm~\ref{dirsolutionalg} with alternate convergence criteria} or \textit{Algorithm~\ref{dirsolutionalg}-ACC} for short.

Figure~\ref{minmaxavg} (left) shows the convergence of the minimum, maximum, and the average of vertex probabilities towards the final fixation probability value for a small graph of 100 nodes. We can observe that the average converges to the final value at a logarithmic rate and much faster than the minimum and maximum vertex probability values. This suggests that while \textit{Algorithm~\ref{dirsolutionalg}-ACC} does not give the same theoretical guarantees as  \textit{Algorithm~\ref{dirsolutionalg}}, it is much preferable for speed since the minimum and maximum vertex probabilities take much longer to converge to the final solution than the average. The fact that the average of the vertex probabilities is much preferable as a fast estimation of fixation probabilities is supported by the logarithmic decrease of the standard deviation of vertex probabilities (see Figure~\ref{stdev}).  Convergences for other and larger graphs are not shown here but are qualitatively similar to the relative convergences shown in the provided graphs. 

\subsection{Speed Comparison to Monte Carlo Simulation}

In order to compare our method's speed compared to the standard Monte Carlo simulation method, we must determine how many iterations our algorithm must be run to find a fixation probability estimate comparable to that of the Monte Carlo approach. Thankfully, as we have seen, we can get a standard error on the fixation probability returned by the Monte Carlo approach as per Equation~\ref{std-err-prop}. While we did not theoretically prove anything about how smoothly fixation  probabilities from our methods approach the final solution, the convergences of the average and standard deviation as shown above strongly suggest that estimates from our method approach the final solution quite gracefully. In fact, in the following experiments, once our method has arrived at a fixation probability estimate within the standard error of simulations, the estimate never again fell outside the window of standard error (although the estimate did not always approach the final estimate monotonically). This is in stark contrast to Monte Carlo simulations, from which estimations can vary greatly before the method has completed enough single runs to achieve a good probability estimate.
 
We generated a number of  randomly weighted and strongly connected directed graphs of various sizes on which we compare our solution method to Monte Carlo approximation of fixation probabilities. The graphs were generated as in our convergence experiments. 
For each graph of a different size, we generated a number of different initial mutant configurations. We found fixation probabilities both using Monte Carlo estimation with 2000 simulation runs and our direct solution method, terminating when we have reached within the standard error of the Monte Carlo estimation.
Since the average vertex probability proved to be such a good fast estimate of the true fixation probability, we used \textit{Algorithm~\ref{dirsolutionalg}-ACC}. 

Figure~\ref{minmaxavg} (right) shows the speedup our solution provides over Monte Carlo simulation. Here speedup is defined as the ratio of the time it takes for simulations to complete over the time it takes our algorithm to find a fixation probability within the standard deviation.   The often extremely low number of iterations needed by our algorithm to find fixation probabilities within the standard error of simulations may prompt the concern that the probabilities fall within this window so soon by mere chance. However, our experiments have shown that the fixation probability estimation found by our algorithm at each iteration approaches the final fixation probability after termination smooothly at a logarithmic rate, asymptotically approaching the true fixation probability. While in this case the fixation probability estimate slightly crosses over the true fixation probability and then slowly approaches it again, none of the fixation probability estimates from our algorithm exited the window of standard error (from simulations) once they entered it.

We can observe from our experiments that computing fixation probabilities using Monte Carlo simulations showed to be a very time-expensive process, highlighting the need for faster solution methods as the one we have presented. Especially for larger graph sizes, the time complexity of our solution to achieve similar results to Monte Carlo simulation has shown to be orders of magnitude smaller than the standard method. 

\subsection{Monitoring the Expected Number of Mutants}

As observed in Section~\ref{exmut-sec}, our method not only allows for the calculation of the fixation probability of a mutant, but also allows us to study how the expected number of mutants change over time.  In this section, we present experimental results exploring the trajectory of the expected number of mutants over time on various undirected/unweighted graphs and under different initial mutant placement conditions.

First, we note that the expected number of mutants (as time approaches infinity) in an unweighted/undirected graph with respect to a single initially infected vertex $i$ can be computed by modifying the result of ~\cite{Broom2010} (for BD updating) to obtain the following.

\begin{equation}
\label{undir-neu-drift-soln}
\EXif = \frac{1}{k_i \langle k^{-1} \rangle}
\end{equation}

Where $\langle k^{-1} \rangle$ is the average inverse of the degree for the graph.  Hence, we can determine whether a node amplifies or suppresses selection by observing if $\EXif$ is greater or less than $1$ respectively: if $k_i < \frac{1}{ \langle k^{-1} \rangle}$ selection is amplified and if $k_i > \frac{1}{\langle k^{-1} \rangle}$ it is suppressed.  We have used our algorithm to compare the trajectory of the expected number of mutants over time when the initial mutant is placed on amplifiers vs. suppressors under different graph topologies and BD updating. We note that similar comparisons can be obtained with our algorithm for the other update rules.  We also note that by Theorem~\ref{bd-mono}, an amplifier for BD (with no bias) will also be an amplifier for the (biased) BD-B and BD-D where $r>1$.

\begin{figure}[th]
  \centering
  \includegraphics[width=1\textwidth]%
    {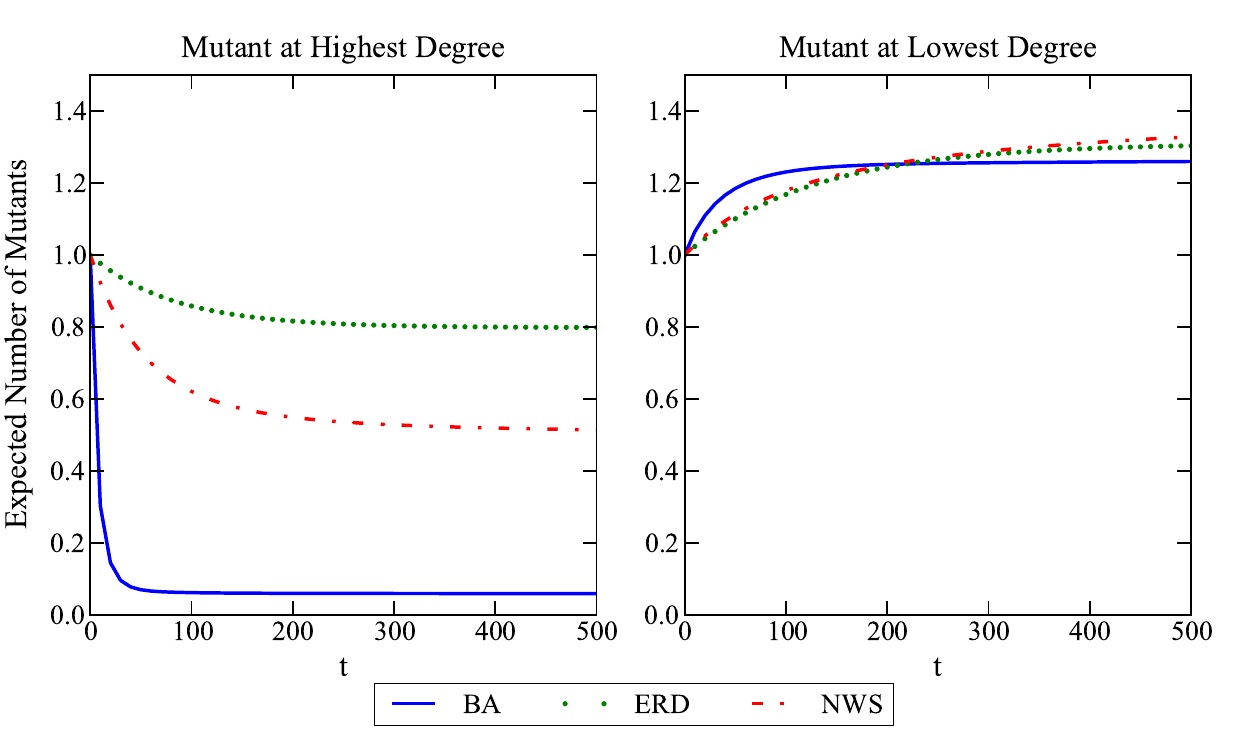}
  \caption{Expected number of mutants over time starting with a single mutant placed on a graph for Barab{\'a}si-Albert preferential attachment (BAR), Erd{\H{o}}s-R{\'e}nyi (ERD), and Newmann-Watts-Strogatz small world (NWS) graphs. Lines are averages over 50 random graphs of each type. In the left graph, mutants are placed at the highest degree nodes, which are suppressors. In the right graph, mutants are placed at lowest degree nodes, which are amplifiers.} \label{ENMtrajs}
\end{figure}

\begin{figure}[th]
  \centering
  \includegraphics[width=1\textwidth]%
    {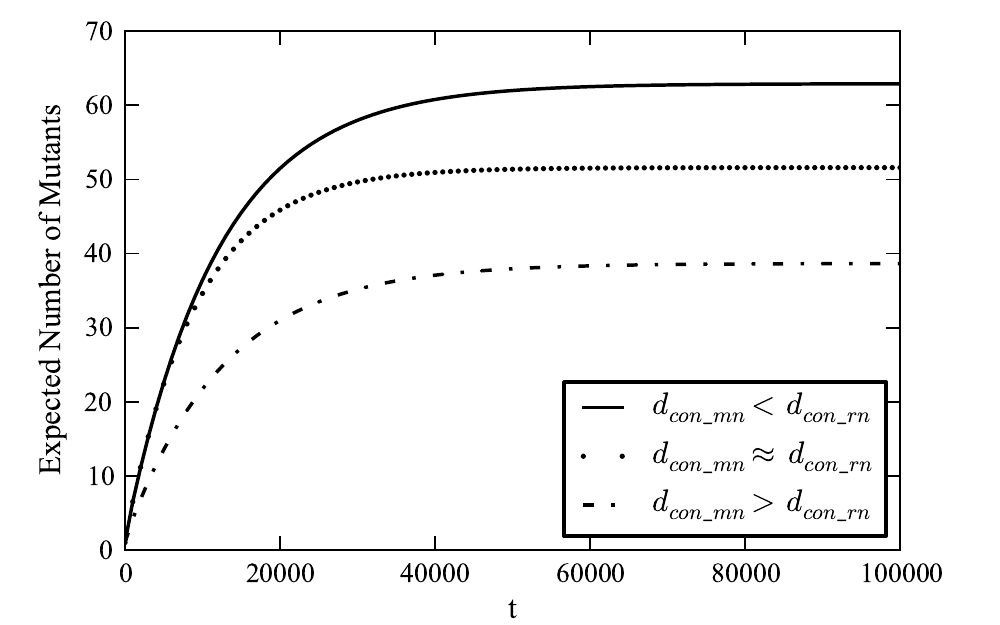}
  \caption{Expected number of mutants over time for an Erd{\H{o}}s-R{\'e}nyi graph of 100 nodes, with an extra muntant node ($mn$) and resident node($rn$) with directed edges to $con\_mn$ and $con\_rn$ respectively. The value that the expected number of mutants converges to depends on the relative degrees of $con\_mn$ and $con\_rn$, as shown in the legend.} 
  \label{ENMspecial}
\end{figure}

Figure~\ref{ENMtrajs} shows the trajectories of the expected number of mutants over time on random \cite{barabasi} preferential attachment (BAR), \cite{erdHos1960evolution} (ER), and  \cite{newman1999renormalization} small world graphs (NWS), each for when the initial mutant is placed on a suppressor (highest degree node of graph) and amplifier (lowest degree node of graph). Graphs are all of equal size at 100 nodes. 
We note that the highest degree nodes are especially strong suppressors on BAR graphs, less so for NWS graphs,  and even less so for ER graphs. This makes sense when one considers the degree distribution of the different graph topologies, which are scale-free or power-law ($P(k) \sim {k}^{-3}$) for BR, roughly Poisson-shaped for NWS, and relatively uniform for ER graphs. For lowest degree amplifiers, the expected number of mutants grows faster early on in Barab{\'a}si-Albert graphs, but it plateaus earlier than and is eventually surpassed by the slower growing expected number of mutants in the Erd{\H{o}}s-R{\'e}nyi, and Newmann-Watts-Strogatz graphs. Such insights into the evolutionary process may be crucial in applications, e.g. when one may be more interested in achieving highest number of mutants in a short amount of time rather than highest number of mutants as $t\to\infty$ or vice versa. 
 
 Finally, thus far we have only considered strongly connected graphs in which the vertex probabilities converge as  $t\to\infty$, but this is not the case for some non-strongly connected graphs. We have thus also investigated the expected number of mutants over time for some simple cases of such graphs. Consider a random graph that is strongly connected, and then have a resident node ($rn$) and mutant node ($mn$) connected with only directed edges into the strongly connected graph. Clearly, the vertex probabilities cannot converge, since $\forall \, t, P_{mn,t}=1$ and $P_{rn,t}=0$. Our experimental results in Figure~\ref{ENMspecial} show however that while the vertex probabilities do not converge, the value for the expected number of mutants given by our algorithm seems to converge. What value the expected number of mutants converges to depends on the relative degrees of the nodes that the mutant node $mn$ and resident node $rn$ connect to.  We shall call these nodes $con\_mn$ and $con\_rn$, respectively. If $k_{con\_mn} \approx k_{con\_rn}$, the expected value of mutants converges at around 50\% of the graph's nodes. If $k_{con\_mn} > k_{con\_rn}$, the expected value of mutants is less than 50\% of the graph's nodes, and conversely, if $k_{con\_mn} < k_{con\_rn}$, it is greater. These results are intuitive because lower degree nodes are better spreaders under BD updating.  These results are also interesting because the expected value converges - even though the graphs are not strongly connected.  By an examination of Equation~\ref{nd-cor2}, this convergence is possible.  However, we have not proven that convergence always occurs.  An interesting direction for future work is to identify under what conditions will the expected number of mutants converges in a non-strongly connected graph.

\subsection{Experimentally Computing the Lower Bound of the Mean Time to Fixation}

We also performed experiments to examine the lower bound on mean time to fixation (discussed in Section~\ref{ttf-sec}) as compared to the average fixation time determined from simulation run.  In doing so, we were able to confirm the lower-bound experimentally.  We were able to use Algorithm~\ref{dirsolutionalg}-ACC to compute the lower bound with a few changes (noted in the supplement).

\begin{figure}[ht]
  \centering
\includegraphics[width=1\textwidth]{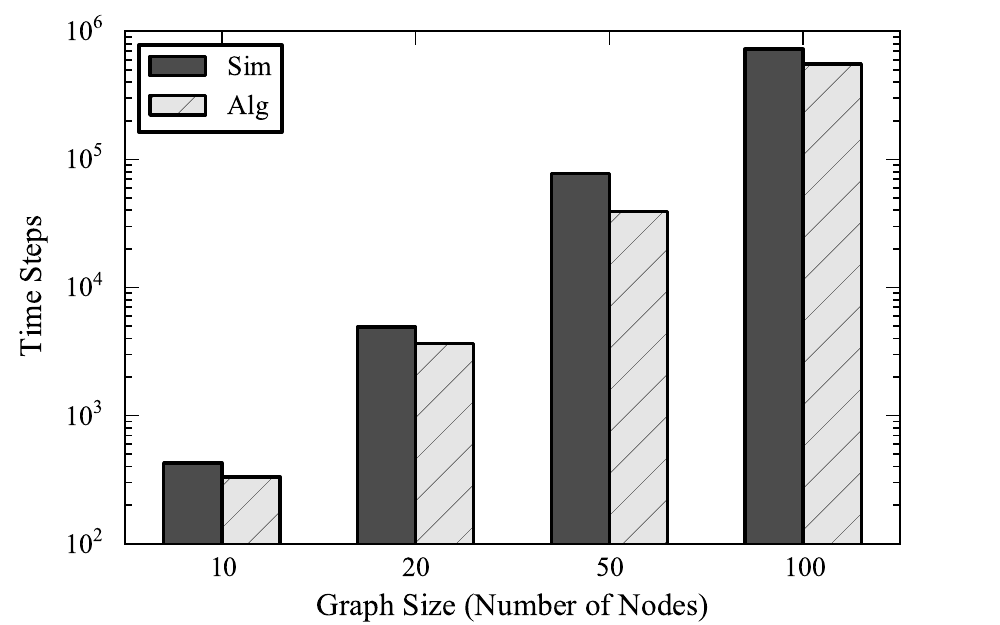}
  \caption{Mean-time-to-fixation comparison between algorithm and simulation.  Note that the y-axis is a logarithmic scale.} \label{fixTime1}
\end{figure}

We generated random (ER) graphs of size $10, 20, 50$ and $100$ nodes, creating five different graphs for each number of nodes.  The graphs were generated as in our convergence experiments, and our comparison to Monte Carlo testing are shown in Figure~\ref{fixTime1} where we demonstrate experimentally that our algorithm produces a lower bound.  Our algorithm was run until the standard deviation of fixation probabilities for all vertices was $2.5 \times 10^{-6}$.  The Monte Carlo simulations were each set at $10,000$ runs. 

\section{Application: Competition Among Neural Axons}
\label{appSec}
In recent work, \cite{turney12} created a model for synaptic competition based on death-birth updating under neutral drift.  They noted that the model aligns well with their empirical observations.  In the model, the graph represents a synaptic junction and the nodes represent sites in the junction.  For every two adjacent sites in the synaptic junction, there is an undirected edge between the corresponding two nodes in the graph. Hence, in- and out- degrees of each node are the same.  Initially, there are $K$ different axon types located in the junction configured in a manner where all sites are initially occupied by one axon type.  At each time step, an axon occupying one of the sites is eliminated - making the site open.  The selection of the axon for elimination (death) is with a uniform probability.  Hence, there is no bias in this model.  Following the elimination of an axon, an adjacent axon grows into the site.  The adjacent axon is selected with a uniform probability of the eliminated axon's neighbors.  Hence, based on the results of this paper, we can provide the following insights into synaptic competition.
\begin{enumerate}
\item\label{neuroFixCalc} After $t$ axons are eliminated,\footnote{Note that the number of axons eliminated corresponds directly to the number of timesteps in the model.} the probability of any site being occupied by an axon of a certain type can be calculated directly by Theorem~\ref{db}.  Even though there are $K$ axon types, this theorem still applies as it only considers the probability of a node being a mutant (resp. a site being a one of the $K$ axon types).
\item Using point~\ref{neuroFixCalc} above, we can determine the expected number of axons of a given type after $t$ axons being eliminated.
\item\label{neuronAdd} After $t$ axons are eliminated, the probability of any set of sites being occupied by a certain axon type is simply the sum of the probabilities of the individual sites being occupied by that axon.  As a result, the fixation probability is additive.
\item\label{neuronFp} Leveraging point~\ref{neuronAdd} above combined with an easy modification of the result of Broom et al.~\cite{Broom2010} for the BD model, the fixation probability of an axon originating at site $i$ is $\frac{k_i}{2\cdot\Theta}$ where $k_i$ is the number of sites adjacent to site $i$ (hence the degree of node $i$ in the corresponding graph) and $\Theta$ is the total number of adjacencies in the synapse (hence, half the number of directed edges in the corresponding graph).
\item Based on item~\ref{neuronFp} above and the results from Section~\ref{exmut-sec}, we can conclude that for a given axon type (let's call it ``axon type A'') occupying a set of sites, that if the average adjacencies of those sites is greater than (resp. less than) the overall average adjacencies for the sites in the entire synaptic junction, then as the number of eliminated axons approaches infinity, we can expect the number of axon type A in the synaptic junction will increase (resp. decrease) in expectation.
\item We can directly apply Theorem~\ref{timeThm} to find a lower-bound on the number of eliminated axons before fixation occurs.
\end{enumerate}
We note that the results stated above are either precise mathematical arguments or calculations that can be found exactly with a deterministic algorithm.  They are not theoretical approximations and do not rely on simulation.  As such is the case, we can make more precise statements about synaptic competition (given the model) and can avoid the variance that accompanies simulation results.  Insights such as these may lead to future biological experiments.

\section{Related Work}
\label{rw-sec}
Evolutionary graph theory was originally introduced in \cite{lieberman_evolutionary_2005}.  Previously, we have compiled a comprehensive review~\citep{Shakarian201266} for a general overview of the work in this exciting new area.

While most work dealing with evolutionary graphs rely on Monte Carlo simulation, there are some good analytical approximations for the undirected/unweighted cased based on the degree of the vertices in question.  Antal et al.~\cite{antal06} use the mean-field approach to create these approximations for the undirected/unweighted case. Broom et al.~\cite{Broom2010} derive an exact analytical result for the undirected/unweighted case in neutral drift, which agrees with the results of Antal et al.  They also show that fixation probability is additive in that case (a result which we extend in this paper using a different proof technique).  However, the results of \cite{masuda09} demonstrate that mean-field approximations break down in the case of weighted, directed graphs. \cite{masuda09a} also studied weighted, directed graphs, but does so by using Monte Carlo simulation.  \cite{rychtar08} derive exact computation of fixation probability through means of linear programming.  However, that approach requires an exponential number of both constraints and variables and is intractable.  The recent work of \cite{voorhees12} introduces a parameter called graph determinacy which measures the degree to which fixation or extinction is determined while starting from a randomly choses initial configuration.  This property is then used to analyze some special cases of evolutionary graphs under birth-death updating.  There has been some work on algorithms for fixation probability calculation that rely on a randomized approach \citep{barbosa2010,egtFpras}.  \cite{barbosa2010} present a heuristic technique for speeding up Monte Carlo simulations by early termination while \cite{egtFpras} present utilize simulation runs in a fully-polynomial randomized approximation scheme.  However, our framework differs in that it does not rely on simulation at all and provides a deterministic result.  Further, our non-randomized approach also allows for additional insights into the evolutionary process - such as monitoring the expected number of mutants as a function of time.  Recently, \cite{bahram11} study the related problem of determining the probability of fixation given a single, randomly placed mutant in the graph where the vertices are ``islands'' and there are many individuals residing on each island in a well-mixed population.  They use quasi-fixed points of ODE's to obtain an approximation of the fixation probability and performed experiments with a maximum of $5$ islands (vertices) containing $50$ individuals each.  This continuous approximation provides the best results when the number of individuals in each island is much larger than the number of islands.  As the problem of this paper can be thought of as a special case where each island has just one individual, it seems unlikely that the approximation of Houchmandzadeh and Vallade's approach will hold here.

Some of the results in this paper were previously presented in conferences by the authors~\citep{shakarianRoos2011,moores12}.  The analysis and experiments concerning the expected number of mutants at a given time, the extension of the framework for other update rules (beyond birth-death), the use of the framework for the case of $r>1$, and the neurology applications are all new results appearing for the first time in this paper.

\section{Conclusion}
\label{conc}

In this paper, we introduced a new approach to deal with problems relating to evolutionary graphs that rely on ``vertex probabilities.'' Our presented analytical method is the first deterministic method to compute fixation probability and provides a number of novel uses and results for EGT problems:

\begin{itemize}
\item Our method can be used to solve for the fixation probability under neutral drift orders of magnitude faster than Monte Carlo simulations, which is currently the presiding employed method in EGT studies. We have extended the method to all of the commonly used update processes in EGT.  The special case of neutral drift is not only of interest in the literature~\citep{Broom2010,masuda09a} but also it has been applied to problems in neurology~\citep{turney12}.
\item While the presented method is currently constrained to the case of neutral drift, we have demonstrated how it can inform cases of non-neutral drift by using it to provide both a lower and upper bound for this case. Combined with our analytical method's speed, this means that it can be used to acquire useful knowledge to guide general EGT studies interested in the case of advantageous mutants.
\item We have shown how our analytical method can be used to calculate a non-trivial lower bound to the mean time to fixation, providing a first step for a general method to computing this and related quantities that is lacking in the current literature.
\item We have shown how our method can be used to calculate deterministically the expected number of mutants, which is useful for applications that require predictions on the number of mutants in the population under a specific finite time horizon. We have also provided results on the expected number of mutants on different common graph topologies, showing differences in the growth trajectories of amplifiers and suppressors on these different topologies.  These results may prove highly significant in the recent application of EGT to distributed systems \citep{jiang12} where the problem of information diffusion is considered among computer systems.  In such a domain, it may be insufficient to guarantee fixation in the limit of time - which may be impractical - but rather to make guarantees on the outcome of the process after a finite amount of time.
\item Finally, we have shown how our method can provide insight when applied to the problem of synaptic competition in neuroscience.
\end{itemize}

Though evolutionary graph theory is still a relatively new research area, it is actively being studied in a variety of disciplines~\citep{lieberman_evolutionary_2005,Shakarian201266,zhang07,sood08,pacheco06,turney12,jiang12}. We believe that more real-world applications will appear as this area gains more popularity. As illustrated by recent work~\citep{turney12,jiang12}, experimental scientists with knowledge of EGT may be more likely to recognize situations where the model may be appropriate. As these cases arise, deterministic methods for addressing issues related to EGT may prove to be highly useful. However, this paper is only a starting point - there are still many important directions for future work. Foremost among such topics are scenarios where the topology of the graph also changes over time or where additional attributes of the nodes/edges in the graph affect the dynamics.

%

\section*{Acknowledgments}
\noindent P.S. is supported by ARO projects 611102B74F and 2GDATXR042 as well as OSD project F1AF262025G001.  P.R. is supported by ONR grant W911NF0810144.  P.S. would like to thank Stephen Turney (Harvard University) for several discussions concerning his work.  The opinions in this paper are those of the authors and do not necessarily reflect the opinions of the funders, the U.S. Military Academy, the U.S. Army, or the U.S. Navy.



\pagebreak
\section*{Supplementary Material}

\section{Notes}
Throughout this supplement, we will use an extended notation.  Fixation probability given initial configuration $C$ is denoted $F_C$.  For vertex $i$ at time $t$, we denote this as  $\Pr(\mutt_i)$.  We will use $\sel_i$ to denote the event that vertex $i$ was selected for reproduction and $\repl_{ij}$ to denote the event of $i$ replacing $j$.  We will often use conditional probabilities.  For example, $\Pr(\mutt_i | \ctz)$ is the probability that $v_i$ is a mutant given the initial set $C$ of mutants.  Throughout this supplement, unless noted otherwise, all of our probabilities will be conditioned on $\ctz$.  We will drop it for ease of notation with the understanding that some set $C$ of $V$ were mutants at $t=0$.  Hence, $\Pr(\mutt_i) = \Pr(\mutt_i | \ctz)$.


\section{Proof of Theorem 1}
\begin{eqnarray*}
\forall i, & \limt \Pr(\mutt_i | \ctz) = F_C
\end{eqnarray*}

\begin{proof}
Consider the following definition property of $\Pr(\mutt_i | \ctz)$

\begin{equation}
\Pr(\mutt_i| \ctz)=\mathop{\sum_{\calv \in 2^V}}_{\textit{s.t. }v_i \in \calv}\Pr(\cvt| \ctz)
\end{equation}
We note that as time approaches infinity, for all $\calv \in 2^V - \emptyset - V$ we have $\Pr(\cvt| \ctz) = 0$.  As $v_i \notin \emptyset$, the statement follows.
\end{proof}


\section{Proof of Theroem 2}

$\Pr(\mutt_i) =$\\
\begin{eqnarray*}
\Pr(\muttl_i)+ \sum_{(v_j,v_i)\in E}w_{ji}\cdot \Pr(\muttl_j)\cdot\Pr(\sel_j|\muttl_j)-w_{ji}\cdot\Pr(\muttl_i)\cdot\Pr(\sel_j | \muttl_i) &&
\end{eqnarray*}

Where $\sel_i$ is true iff $v_i$ is selected for reproduction at time $t$.

\begin{proof}
Note we use the variable $\repl_ji$ is true iff $v_j$ replaces $v_i$ at time $t$.\\

\noindent CLAIM 1:
\begin{eqnarray*}
\Pr(\mutt_i) &=& \Pr(\muttl_i \wedge \mathop{\bigwedge}_{(v_j,v_i)\in E}\neg \sel_j)+ \sum_{(v_j,v_i)\in E}\Pr(\sel_j \wedge \repl_{ji} \wedge \muttl_j)+\\
&& \sum_{(v_j,v_i)\in E}\Pr(\sel_j \wedge \neg\repl_{ji} \wedge \muttl_i)
\end{eqnarray*}
This is shown by a simple examination of exhaustive and mutually exclusive events based on the original model of \cite{lieberman_evolutionary_2005}.\\

\noindent CLAIM 2:
\begin{eqnarray*}
\Pr(\muttl_i \wedge \mathop{\bigwedge}_{(v_j,v_i)\in E}\neg \sel_j) &=& \Pr(\muttl_i)\cdot \left(1-\sum_{(v_j,v_i)\in E}\Pr(\sel_j | \muttl_i)\right)
\end{eqnarray*}

\noindent(Proof of claim 2) 
By exhaustive and mutual exclusive events, we have the following.
\begin{eqnarray*}
\Pr(\muttl_i \wedge \mathop{\bigwedge}_{(v_j,v_i)\in E}\neg \sel_j) &=& \Pr(\muttl_i) - \sum_{(v_j,v_i)\in E}\Pr(\sel_j \wedge \muttl_i)
\end{eqnarray*}
By the definition of conditional probability, we have the following
\begin{eqnarray*}
\Pr(\muttl_i \wedge \mathop{\bigwedge}_{(v_j,v_i)\in E}\neg \sel_j) &=& \Pr(\muttl_i) - \sum_{(v_j,v_i)\in E}\left( \Pr(\sel_j | \muttl_i) \cdot \Pr(\muttl_i) \right)\\
&=& \Pr(\muttl_i) - \Pr(\muttl_i)\cdot\sum_{(v_j,v_i)\in E}\Pr(\sel_j | \muttl_i)\\
&=& \Pr(\muttl_i) \left( 1- \sum_{(v_j,v_i)\in E}\Pr(\sel_j | \muttl_i)\right)
\end{eqnarray*}
The claim immediately follows.\\

\noindent CLAIM 3: For all edges $(v_j,v_i)$, we have the following.
\begin{eqnarray*}
\Pr(\sel_j \wedge \repl_{ji} \wedge \muttl_j) &=& w_{ji}\cdot \Pr(\muttl_j)\cdot\Pr(\sel_j|\muttl_j)
\end{eqnarray*}
\noindent (Proof of claim 3)
The following is a direct application of the definition of conditional probability.
\begin{eqnarray*}
\Pr(\sel_j \wedge \repl_{ji} \wedge \muttl_j) &=& \Pr(\repl_{ji} \wedge \muttl_j | \sel_j)\cdot\Pr(\sel_j)
\end{eqnarray*}
From our model, we note that given even $\sel_j$, the fitness of the nodes is not considered in determining if the event associated with $\repl_{ji}$ is to occur.  Hence, it follows that $\muttl_j$ is independent of $\repl_{ji}$ given $\sel_j$.  As $\Pr(\repl_{ji}|\sel_j)=w_{ji}$, we have the following.
\begin{eqnarray*}
\Pr(\sel_j \wedge \repl_{ji} \wedge \muttl_j) &=& \Pr(\repl_{ji} | \sel_j)\cdot\Pr(\muttl_j | \sel_j)\cdot\Pr(\sel_j)\\
&=& w_{ji} \cdot\Pr(\muttl_j | \sel_j)\cdot\Pr(\sel_j)
\end{eqnarray*}
By Bayes Theorem, and that the model causes $\forall i \Pr(\sel_i)>0$, we have the following.
\begin{eqnarray*}
\Pr(\sel_j \wedge \repl_{ji} \wedge \muttl_j) &=& w_{ji} \cdot\Pr(\sel_j | \muttl_j )\cdot\frac{\Pr(\muttl_j)}{\Pr(\sel_j)}\cdot\Pr(\sel_j)\\
&=& w_{ji}\cdot \Pr(\muttl_j)\cdot\Pr(\sel_j|\muttl_j)
\end{eqnarray*}
The claim follows immediately.\\

\noindent CLAIM 4: For all edges $(v_j,v_i)$, we have the following.
\begin{eqnarray*}
\Pr(\sel_j \wedge \neg\repl_{ji} \wedge \muttl_i) &=& (1-w_{ji})\cdot \Pr(\muttl_i)\cdot\Pr(\sel_j|\muttl_i)
\end{eqnarray*}
\noindent (Proof of claim 4) This mirrors claim 3.\\

\noindent (Proof of theorem) From claims 1-4, we have the following.
\begin{eqnarray}
\Pr(\mutt_i) &=& \Pr(\muttl_i)\cdot \left(1-\sum_{(v_j,v_i)\in E}\Pr(\sel_j | \muttl_i)\right)+\\
&& \sum_{(v_j,v_i)\in E}\left(w_{ji}\cdot \Pr(\muttl_j)\cdot\Pr(\sel_j|\muttl_j) \right)+\\
&& \sum_{(v_j,v_i)\in E}\left((1-w_{ji})\cdot \Pr(\muttl_i)\cdot\Pr(\sel_j|\muttl_i) \right)
\end{eqnarray}
Which, after re-arranging some terms, gives us the statement of the theorem.
\end{proof}


\section{Proof of Theorem 3}

When $r=1$, if for some time $t$,  $\forall i$, the value $\Pr(\mutt_i)$ is the same, then $\Pr(\mutt_i)=F_C$.\\

\noindent\textit{\textbf{Proof Sketch.}  Consider $\Pr(\mutt_i) =\Pr(\muttl_i)+ \frac{1}{N}\sum_{(v_j,v_i)\in E}w_{ji}\cdot (\Pr(\muttl_j)-\Pr(\muttl_i))$ when for $t-1$, $\forall i,j$ we have $\Pr(\muttl_j)=\Pr(\muttl_i)$.  Clearly, in this case, the value for $\Pr(\mutt_i) =\Pr(\muttl_i)$.  As the probabilities of all vertices was the same at $t-1$, they remain so at $t$.  Therefore, in this case, $\limt\Pr(\mutt_i) = \Pr(\mutt_i)$. $\blacksquare$}\\


\section{Proof of Inequality 4}
For any time $t$, under neutral drift ($r=1$), 
\begin{equation*}
\min_i \Pr(\mutt_i) \leq F_C \leq \max_i \Pr(\mutt_i)
\end{equation*}

\begin{proof}
\noindent PART 1: For any time $t$, under neutral drift ($r=1$), $F_C \leq \max_i \Pr(\mutt_i)$.\\
We show that for each time step $t$, $\max_i \Pr(\muttl_i)\geq \max_i \Pr(\mutt_i)$.  Hence, by showing that, for any time $t'$, we have $\max_i \Pr(\muttp_i)\geq \lim_{t \rightarrow \infty}\max_i \Pr(\mutt_i)$ which by allows us to apply Theorem 1 and obtain the statement of this theorem.  Suppose BWOC that at time $t$ we have $\max_\ell \Pr(\muttl_\ell)< \max_i \Pr(\mutt_i)$.  Then we have:
\begin{eqnarray*}
\max_\ell \Pr(\muttl_\ell) &<& \frac{1}{N}\sum_{(v_j,v_i)\in E}w_{ji}\cdot \left(\Pr(\muttl_j)-\Pr(\muttl_i)\right) \\
&&+\Pr(\muttl_i)\\
&\leq& \frac{1}{N}\sum_{(v_j,v_i)\in E}w_{ji}  \cdot \left(\max_\ell \Pr(\muttl_\ell)-\Pr(\muttl_i)\right)\\
&&+\Pr(\muttl_i)\\
&=& \frac{\sum_{(v_j,v_i)\in E}w_{ji}}{N} \left(\max_\ell \Pr(\muttl_\ell)-\Pr(\muttl_i)\right)\\
&&+\Pr(\muttl_i)\\
\max_\ell \Pr(\muttl_\ell)(1 -\frac{\sum_{(v_j,v_i)\in E}w_{ji}}{N})&<&  \Pr(\muttl_i)(1 -\frac{\sum_{(v_j,v_i)\in E}w_{ji}}{N})\\
\max_\ell \Pr(\muttl_\ell)&<&  \Pr(\muttl_i)
\end{eqnarray*}
Which is clearly a contradiction and completes this part of the proof.\\
\noindent PART 2: For any time $t$, under neutral drift ($r=1$), $F_C \geq \min_i \Pr(\mutt_i)$.\\
We show that for each time step $t$, $\min_i \Pr(\muttl_i)\leq \min_i \Pr(\mutt_i)$.  Hence, by showing that, for any time $t'$, we have $\min_i \Pr(\muttp_i)\leq \lim_{t \rightarrow \infty}\max_i \Pr(\mutt_i)$ which by allows us to apply Theorem 1 and obtain the statement of this theorerm.  Suppose BWOC that at time $t$ we have $\min_\ell \Pr(\muttl_\ell)> \min_i \Pr(\mutt_i)$.  Then we have:
\begin{eqnarray*}
\min_\ell \Pr(\muttl_\ell) &>& \frac{1}{N}\sum_{(v_j,v_i)\in E}w_{ji}\cdot \left(\Pr(\muttl_j)-\Pr(\muttl_i)\right) \\
&&+\Pr(\muttl_i)\\
&\geq& \frac{1}{N}\sum_{(v_j,v_i)\in E}w_{ji}  \cdot \left(\min_\ell \Pr(\muttl_\ell)-\Pr(\muttl_i)\right)\\
&&+\Pr(\muttl_i)\\
&=& \frac{\sum_{(v_j,v_i)\in E}w_{ji}}{N} \left(\min_\ell \Pr(\muttl_\ell)-\Pr(\muttl_i)\right)\\
&&+\Pr(\muttl_i)\\
\min_\ell \Pr(\muttl_\ell)(1 -\frac{\sum_{(v_j,v_i)\in E}w_{ji}}{N})&>&  \Pr(\muttl_i)(1 -\frac{\sum_{(v_j,v_i)\in E}w_{ji}}{N})\\
\min_\ell \Pr(\muttl_\ell)&>&  \Pr(\muttl_i)
\end{eqnarray*}
Which is clearly a contradiction and completes this part of the proof.
\end{proof}

\section{Proof of Theorem Theorem 4}
When $r=1$ for disjoint sets $C,D \subseteq V$, $F_C+F_D = F_{C \cup D}$.

\begin{proof}
Consider some time $t$ and vertex $v_i$.  Clearly, by Corollary 1, $\Pr(\mutt_i)$ can be expressed as a linear combination of the form $\sum_{v_j \in V}(C_j \cdot \Pr(\mutz_j))$ where $C_j$ is a coefficient.  We note that these coefficients are the same regardless of the initial configuration of mutants that $\mutt_i$ is conditioned on.  Hence, $\Pr(\mutt_i | C^{(0)})$ is this positive function with $\Pr(\mutz_j)=1$ if $v_j \in C$ and $0$ otherwise (see Proposition 3).  Hence, for disjoint $C,D$, for any $v_i \in V$, we have $\Pr(\mutt_i | C^{(0)})+\Pr(\mutt_i | D^{(0)})=\Pr(\mutt_i | (C\cup D)^{(0)})$.  The statement follows.
\end{proof}


\section{Proof of Equation 9}
\begin{eqnarray*}
\Pr(\mutt_i)=\left( 1-\frac{1}{N}\right) \cdot \Pr(\muttl_i)+\frac{1}{N \cdot \dini}\mathop{\sum}_{(v_j,v_i)\in E} \Pr(\muttl_j)
\end{eqnarray*}
Under death-birth dynamics with neutral drift ($r=1$).

\begin{proof}
$\deatht_i$ and $\birtht_i$ are random variables associated with birth and death events for vertex $v_i$.\\
\noindent CLAIM 1: $\Pr(\mutt_i)=\Pr(\muttl_i \wedge \neg \deatht_i) + \sum_{(v_j,v_i)\in E} \Pr(\deatht_i \wedge \birtht_j \wedge \muttl_j)$\\
Follows directly form exhaustive and mutually exclusive events.\\
\noindent CLAIM 2: $\Pr(\muttl_i \wedge \neg \deatht_i) = \left( 1-\frac{1}{N}\right) \cdot \Pr(\muttl_i)$\\
By the definition of conditional probabilities, we have $\Pr(\neg \deatht_i | \muttl_i) \cdot \Pr(\muttl_i)$.  Also, we know the probability of a given node dying is always $1/N$.  Hence, $\Pr(\neg \deatht_i | \muttl_i) = \Pr(\neg \deatht_i) =  1-\frac{1}{N}$ and the claim follows.\\
\noindent CLAIM 3: For any $(v_j,v_i)\in E$, we have\\ $\Pr(\deatht_i \wedge \birtht_j \wedge \muttl_j)=\frac{1}{N \cdot \dini} \cdot \Pr(\muttl_j)$\\
As both birth and death events occur independent of any node being a mutant at the previous time step, the definition of conditional probabilities gives us the following:
\begin{eqnarray}
\Pr(\deatht_i \wedge \birtht_j \wedge \muttl_j)&=&\Pr(\deatht_i \wedge \birtht_j)\cdot\Pr(\muttl_j)\\
&=&\Pr(\birtht_j | \deatht_i) \cdot \Pr(\deatht_i) \cdot\Pr(\muttl_j)
\end{eqnarray}
From the model, we have the following:
\begin{eqnarray}
\Pr(\birtht_j | \deatht_i)&=& 1/\dini\\
\Pr(\deatht_i)&=& 1/N
\end{eqnarray}
Hence, the claim follows.
$\blacksquare$\end{proof}


\section{Proof of Equation 10}
\begin{eqnarray*}
\Pr(\mutt_i)=\left( 1-\frac{\dini}{|E|} \right) \cdot \Pr(\muttl_i)+\frac{1}{|E|}\mathop{\sum}_{(v_j,v_i) \in E}\Pr(\muttl_j)
\end{eqnarray*}
Under link dynamics with neutral drift ($r=1$).

\begin{proof}
Here $\sel_{ij}$ is the random variable associated with the selection of edge $(v_i,v_j)$.\\
\noindent CLAIM 1: $\Pr(\mutt_i)=\Pr(\muttl_i \wedge \bigwedge_{(v_j,v_i) \in E}\neg \sel_{ji})+\sum_{(v_j,v_i) \in E}\Pr(\sel_{ji}\wedge \muttl_j)$\\
Follows directly form exhaustive and mutually exclusive events.\\
\noindent CLAIM 2: $\Pr(\bigwedge_{(v_j,v_i) \in E}\neg \sel_{ji})=1-\frac{\dini}{|E|}$\\
Clearly, we have $\Pr(\bigwedge_{(v_j,v_i) \in E}\neg \sel_{ji})=\Pr(\bigvee_{\{(v_\beta,v_\alpha) \in E | \beta \neq i\}}\sel_{\beta\alpha})$.  As there are $\dini$ incoming edges to $v_i$, we know that $\Pr(\bigvee_{\{(v_\beta,v_\alpha) \in E | \beta \neq i\}}\sel_{\beta\alpha})=1-\frac{\dini}{|E|}$, giving us the claim.\\
\noindent CLAIM 3: $\Pr(\muttl_i \wedge \bigwedge_{(v_j,v_i) \in E}\neg \sel_{ji})=\left(1-\frac{\dini}{|E|} \right) \cdot \Pr(\muttl_i)$\\
For any $\alpha,\beta$, the random variable $\sel_{\alpha\beta}$ is independent from $\muttl_i$.  Hence, the claim immediately follows from this fact and claim 2.\\
\noindent CLAIM 4: $\Pr(\sel_{ji}\wedge \muttl_j)=\frac{1}{|E|}\cdot\Pr(\muttl_j)$\\
As, by the definition of the model, $\Pr(\sel_{ji} | \muttl_j)=\Pr(\sel_{ji})=\frac{1}{|E|}$, the claim follows directly form the definition of conditional probabilities.
$\blacksquare$\end{proof}


\section{Proof of Theorem 5}
For a given set $C$, let $\fPo(C)$ be the fixation probability under neutral drift and $\fPr(C)$ be the fixation probability calculated using a mutant fitness $r>1$.  Then, under BD-B, BD-D, DB-B, DB-D, or LD dynamics, $\fPo(C) \leq \fPr(C)$.

\begin{proof}
First, some notation.
\begin{itemize}
\item{We define an \textbf{interpretation}, $\I : 2^V \rightarrow [0,1]$ as probability distribution over mutant configurations.  Hence, for some $\I$ we have $\sum_{\calv \in 2^V}\I(\calv)=1$.}
\item{Next, we define a transition function that maps configurations of mutants to probabilities, $\chi : 2^V \rightarrow [0,1]$ where for any $C \in 2^V$, $\sum_{C'\in 2^V}\chi(C,C')=1$.  We will use $\incchi$ and $\decchi$ to indicate if the transition is made with a mutant being selected for birth ($\incchi$) or resident ($\decchi$).  Hence, for some $C \in V$ and $v \notin C$, $\decchi(C,C\cup \{v \})=0$ and $\incchi(C\cup \{v \},C)=0$.  Hence, for all $C \in 2^V$, $\sum_{C'\in 2^V}( \incchi(C,C')+\decchi(C,C'))=1$.}
\item{If the transitioon function is based on birth-death and computed with some $r>1$, then we will write it as $\incchir,\decchir$ respectively.  If computed with $r=1$, then we write $\incchin,\decchin$ respectively.}
\item{For some $C \in 2^C$, let $\inc(C)$ be the set of all elements $D \in 2^V$ s.t. $|D| \geq |C|$ and $\incchi(C,D)>0$.}
\item{For some $C \in 2^C$, let $\dec(C)$ be the set of all elements $D \in 2^V$ s.t. $|D| \leq |C|$ and $\decchi(C,D)>0$.}
\item{Given set $C \subseteq V$, we will use $\fPr_{C}$ to denot the probability of fixation given initial set of mutants $C$ where the value $r$ is used to calculate all transition probabilities.}
\end{itemize}

\noindent CLAIM 1: If a some time period, the probability distribution over mutant configurations is $\I$, the fixation probability is $\sum_{C \in 2^V}\I(C)\cdot \fPr_{C}$.\\

Clearly, for any time $t$, $\fPr_{C} = \mathit{lim}_{i\rightarrow \infty}\Pr(V^{(i)}|C^{(t)})$.  Under the assumption that there exists some tim $\omega$ s.t. fixation is reached, we have:
\begin{eqnarray*}
\fPr_{C} &=& \Pr(V^{(\omega)}|C^{(t)})\\
&=&\frac{\Pr(V^{(\omega)} \wedge C^{(t)})}{\Pr(C^{(t)})}
\end{eqnarray*}
Hence, $\fPr_{C} \cdot \Pr(C^{(t)}) = \Pr(V^{(\omega)} \wedge C^{(t)})$.  The statement then follows by the summation of exhaustive and mutually exclusive events.\\

\noindent CLAIM 2: If a some time period $t$, the probability distribution over mutant configurations is $\I$, and the transition functions used to reach the next time step are $\incchi,\decchi$, then the probability of being in some mutant configuration $C$ at time $t+1$ is given by $\sum_{D \in 2^V}\I(D)\cdot(\incchi(D,C)+\decchi(D,C))$.\\

Follows directly from the rules of dynamics.\\

\noindent CLAIM 3: If a some time period $t$, the probability distribution over mutant configurations is $\I$, mutant fitness $r$, and the transition functions used to reach the next time step are $\incchir,\decchir$, and all subsequent transitions are computed using the same dynamics with neutral drift, then the fixation probability is:
\begin{eqnarray*}
\calp(I,r)&=&\sum_{C \in 2^V}\I(C)\cdot\left(\sum_{D \in \inc(C)}(\incchir(C,D)\cdot \fPo_{D})+\sum_{D \in \dec(C)}(\decchir(C,D)\cdot \fPo_{D}))\right)
\end{eqnarray*}

Follows directly from claims 1-2.\\

\noindent CLAIM 4: Under BD-B, BD-D, DB-B, DB-D, or LD dynamics, for some $r \leq r'$, for all $C,D \in 2^V$, we have $\incchir(C,D) \leq \incchirp(C,D)$ and $\decchir(C,D) \geq \decchirp(C,D)$.\\

\noindent CLAIM 4a: For some $r \leq r'$, for all $C,D \in 2^V$, we have $\incchir(C,D) \leq \incchirp(C,D)$.\\
Let $\{v_j\} = D-C$.  For each vertex $v_i$, $f_i =1$ if $v_i \notin C$ (a resident) and $f_i=r$ if $v_i \in C$ (a mutant).  When $D \equiv C$, the following are all summed over the set $\{v_j \in C | \exists v_i\in C \wedge (v_i,v_j)\in E\}$.
\begin{itemize}
\item{Under BD-B, 
\begin{eqnarray*}
\incchir(C,D) &=& \mathop{\sum_{v_i \in C |}}_{(v_i,v_j)\in E}\frac{r\cdot w_{ij}}{r\cdot|C|+N-|C|}
\end{eqnarray*}
}
\item{Under BD-D,
\begin{eqnarray*}
\mathop{\sum_{v_i \in C |}}_{(v_i,v_j)\in E}\frac{w_{ij}}{N\cdot\sum_{v_q | (v_i,v_q)\in E}w_{iq}\cdot f_q^{-1}}
\end{eqnarray*}
}
\item{Under DB-B,
\begin{eqnarray*}
\incchir(C,D) &=& \mathop{\sum_{v_i \in C|}}_{(v_i,v_j)\in E}\frac{w_{ij}\cdot r}{N \cdot \sum_{v_q | (v_q,v_j)\in E}w_{qj}\cdot f_q}
\end{eqnarray*}
}
\item{Under DB-D,
\begin{eqnarray*}
\incchir(C,D) &=& \mathop{\sum_{v_i \in C |}}_{(v_i,v_j)\in E}\frac{w_{ij}}{\sum_{v_q \in V}f_q^{-1}}
\end{eqnarray*}
}
\item{Under LD,
\begin{eqnarray*}
\incchir(C,D) &=& \sum_{v_i\in C | (v_i,v_j)\in E}\frac{w_{ij}\cdot r}{\sum_{v_q,v_{\ell} | (v_q,v_{\ell})\in E}w_{q\ell}\cdot f_q}
\end{eqnarray*}
}
\end{itemize}
By simple algebraic manipulation, for each of these, when all values other than $r$ are fixed, they increase as $r$ increases.\\

\noindent CLAIM 4b: For some $r \leq r'$, for all $C,D \in 2^V$, we have $\decchir(C,D) \geq \decchirp(C,D)$.
Let $\{v_j\} = C-D$.  For each vertex $v_i$, $f_i =1$ if $v_i \notin C$ (a resident) and $f_i=r$ if $v_i \in C$ (a mutant).  When $D \equiv C$, the following are all summed over the set $\{v_j \in V-C | \exists v_i\in V-C \wedge (v_i,v_j)\in E\}$.
\begin{itemize}
\item{Under BD-B, 
\begin{eqnarray*}
\incchir(C,D) &=& \mathop{\sum_{v_i \in V-C |}}_{(v_i,v_j)\in E}\frac{w_{ij}}{r\cdot|C|+N-|C|}
\end{eqnarray*}
}
\item{Under BD-D,
\begin{eqnarray*}
\mathop{\sum_{v_i \in V-C |}}_{(v_i,v_j)\in E}\frac{w_{ij}\cdot r^{-1}}{N\cdot\sum_{v_q | (v_i,v_q)\in E}w_{iq}\cdot f_q^{-1}}
\end{eqnarray*}
}
\item{Under DB-B,
\begin{eqnarray*}
\incchir(C,D) &=& \mathop{\sum_{v_i \in V-C|}}_{(v_i,v_j)\in E}\frac{w_{ij}}{N \cdot \sum_{v_q | (v_q,v_j)\in E}w_{qj}\cdot f_q}
\end{eqnarray*}
}
\item{Under DB-D,
\begin{eqnarray*}
\incchir(C,D) &=& \mathop{\sum_{v_i \in V-C |}}_{(v_i,v_j)\in E}\frac{w_{ij}\cdot r^{-1}}{\sum_{v_q \in V}f_q^{-1}}
\end{eqnarray*}
}
\item{Under LD,
\begin{eqnarray*}
\incchir(C,D) &=& \sum_{v_i\in V-C | (v_i,v_j)\in E}\frac{w_{ij}}{\sum_{v_q,v_{\ell} | (v_q,v_{\ell})\in E}w_{q\ell}\cdot f_q}
\end{eqnarray*}
}
\end{itemize}
By simple algebraic manipulation, for each of these, when all values other than $r$ are fixed, they decrease as $r$ increases.\\

\noindent CLAIM 5: Given some $C \in 2^V$, for all pairs $D,D'$ where $D \in \inc(C)$ and $D' \in \dec(C)$, we have $\fPo_D \geq \fPo_{D'}$.\\

Follows directly from Theorem 5.\\

\noindent CLAIM 6: Given interpretation $\I$, under BD-B, BD-D, DB-B, DB-D, or LD dynamics, for some $r >1$, $\calp(\I,r)\geq \calp(\I,0)$.\\
Let us consider some set $C$ from the outermsot summation in the computation of $\calp(\I,r)$.  Suppose, BWOC, there exists some $C \in 2^V$ s.t.:
\begin{scriptsize}
\begin{eqnarray*}
\sum_{D \in \inc(C)}(\incchir(C,D)\cdot \fPo_{D})+\sum_{D \in \dec(C)}(\decchir(C,D)\cdot \fPo_{D})&<&\sum_{D \in \inc(C)}(\incchio(C,D)\cdot \fPo_{D})+\sum_{D \in \dec(C)}(\decchio(C,D)\cdot \fPo_{D})
\end{eqnarray*}
\end{scriptsize}
This give us:
\begin{scriptsize}
\begin{eqnarray*}
\sum_{D \in \inc(C)}(\incchir(C,D)\cdot \fPo_{D})-\sum_{D \in \inc(C)}(\incchio(C,D)\cdot \fPo_{D})&<&\sum_{D \in \dec(C)}(\decchio(C,D)\cdot \fPo_{D})-\sum_{D \in \dec(C)}(\decchir(C,D)\cdot \fPo_{D})
\end{eqnarray*}
\end{scriptsize}
Let $F_{sm} =\mathit{inf}\{\fPo_D | D \in \inc(C) \}$ and $F_{lg} =\mathit{sup}\{\fPo_D | D \in \dec(C) \}$, this give us:
\begin{scriptsize}
\begin{eqnarray*}
F_{sm}\sum_{D \in \inc(C)}(\incchir(C,D)-\incchio(C,D))&<&F_{lg}\sum_{D \in \dec(C)}(\decchio(C,D)-\decchir(C,D))
\end{eqnarray*}
\end{scriptsize}
Consider ther following:
\begin{scriptsize}
\begin{eqnarray*}
\sum_{D \in \inc(C)}\incchir(C,D)+\sum_{D \in \dec(C)}\decchir(C,D)&=&\sum_{D \in \inc(C)}\incchio(C,D)+\sum_{D \in \dec(C)}\decchio(C,D)\\
\sum_{D \in \inc(C)}\incchir(C,D)-\sum_{D \in \inc(C)}\incchio(C,D)&=&\sum_{D \in \dec(C)}\decchio(C,D)-\sum_{D \in \dec(C)}\decchir(C,D)
\end{eqnarray*}
\end{scriptsize}
Note that by claim 4, both sides of the above equation are positive numbers.  Hence, we have $F_{sm}<F_{lg}$, which contradicts claim 5.\\

\noindent PROOF OF THEOREM:  Let $\calp^{(1)}(\I,r)= \calp(\I,r)$ and $\calp^{(i+1)}(\I,r)= \calp(\calp^{(i)}(\I,r),r)$.  By claim 6, for any $i$, $\calp^{(i+1)}(\I,r)\geq \calp(\calp^{(i)}(\I,r),r)$.  Consider an interpretation $\I$ that describes the initial probability distribution over mutant configurations.  The fixation probability under neutral drift is $\calp(\I,1)$.  For some value $r>1$, the fixation probaiblity is $\mathit{lim}_{i \rightarrow \infty}\calp^{(i)}(\I,r)$.  Clearly, $\mathit{lim}_{i \rightarrow \infty}\calp^{(i)}(\I,r)\geq \calp(\I,1)$.
\end{proof}

\section{Proof of Theorem 6}

\begin{eqnarray*}
t_C = \frac{1}{F_C}\sum_{t=1}^{\infty}t\cdot (\PCt-\PCtl)
\end{eqnarray*}

\begin{proof}
This proof was first presented in \cite{antalTime06}.  The mean time to fixation is described as the expected time to fixation given that the process fixates.  Let $\PCt$ be the probability that fixation is reached in exactly $t$ time-steps or less.  Hence, the probability of reaching fixation in exactly $t$ time steps conditioned on the process reaching fixation is $(\PCt-\PCtl)/F_C$.  The remainder of the theorem follows from the definition of an expected value.
\end{proof}


\section{Proof of Theorem 7}
We introduce two pieces of notation $t_{fix}, t_{convg}$.  We define $t_{fix}$ as a time s.t. $t_C = \frac{1}{F_C}\sum_{t=1}^{t_{fix}}t\cdot (\PCt-\PCtl)$. and $t_{convg}$ s.t. $\forall i$, $\Pr(\muttcv_i)=F_C$.  While in reality, both of these values could be infinite, we note that the relationship $\infty \geq t_{fix} \geq t_{convg}$ holds and that using a lower value for $t_{fix}$ and/or $t_{convg}$ will still ensure we have a lower bound.

\begin{eqnarray*}
\frac{1}{F_C}\sum_{t=1}^{\infty}t\cdot (\prmnt-\prmntl)\leq \frac{1}{F_C}\sum_{t=1}^{\infty}t\cdot (\PCt-\PCtl)
\end{eqnarray*}
Where $\prmnt = \min_i \Pr(\mutt_i)$.

\begin{proof}
First, we have the following.
\begin{equation}
\sum_{t=1}^{\tcv}t\cdot (\prmnt-\prmntl)\leq \sum_{t=1}^{\tfx}t\cdot (\PCt-\PCtl)
\end{equation}
For any time $t$, let $\calpt = \PCt-\PCtl$ and $\prdmnt=\prmnt-\prmntl$. As for each time $t$, we have $\prmnt\geq \PCt$, we can define $\calpttp$ as the portion of $\calpt$ accounted for in $\prdmntp$.  This results in having $\calpttp = 0$ whenever $t'>t$ or $t>\tcv$ as well as the following:
\begin{eqnarray}
\calpt &=& \sum_{t'=1}^{t}\calpttp\\
\prdmnt &=& \sum_{t'=t}^{\tfx}\calpptt\\
\sum_{t=1}^{\tcv}t\cdot (\prmnt-\prmntl) &=& \sum_{t=1}^{\tcv}\sum_{t'=t}^{\tfx}t\calpptt\\
 &=& \sum_{t=1}^{\tcv}\sum_{t'=1}^{\tfx}t\calpptt\\
 &=& \sum_{t'=1}^{\tfx}\sum_{t=1}^{\tcv}t\calpptt\\
 &=& \sum_{t'=1}^{\tfx}\sum_{t=1}^{t'}t\calpptt
 \end{eqnarray}
We also note that the following is true:
\begin{eqnarray}
\sum_{t=1}^{\tfx}t\cdot (\PCt-\PCtl) &=&\sum_{t'=1}^{\tfx}t' \calptp \\
&=& \sum_{t'=1}^{\tfx}t'  \sum_{t=1}^{t'}\calpptt\\
&\geq&\sum_{t'=1}^{\tfx}\sum_{t=1}^{t'}t\calpptt\\
&=&\sum_{t=1}^{\tcv}t\cdot (\prmnt-\prmntl)
 \end{eqnarray}
 Which concludes the proof.
\end{proof}

\section{Materials and Methods}
Except for the experiments dealing with time to fixation/extinction, all algorithms were implemented in Python and run on a 2.33GHz  Intel Xeon CPU.  The time-to-fixation experiments were run on a machine equipped with an Intel Core i7 M620 processor running at $2.67$ GHz with $4$ GB RAM.

All graphs in the experiments were generated using the Python NetworkX package~\cite{hagberg-2008-exploring}. Parameters used for the experiments concerning the expected number of mutants were m=1 for BA, p = 0.5 for ER, and $k=2$ and $p=0.5$ for NWS graph generator functions.

We modified Algorithm 1-ACC based on the results on mean time to fixation as follows:
\begin{itemize}
\item Before line 14, insert: \textbf{t += 1; Sum += t*(min(p)-min(q))}
\item Replace line 17 with: \textbf{return = Sum/average(p)}, where average(q) is the algorithm's best estimate for P$_c$ at termination.
\end{itemize}
\label{lastpage}

\end{document}